\documentclass[twocolumn,prl,showpacs,amsmath,amssymb]{revtex4}
\usepackage{amsmath}
\usepackage{dcolumn}
\usepackage[dvips]{graphicx,color}
\usepackage{wrapfig}

\newcommand{\bra}{\left\langle}
\newcommand{\ket}{\right\rangle}

\newcommand{\der}[2]{\dfrac{d #1}{d  #2}}
\newcommand{\bv}[1]{{\boldsymbol #1}}

\newcommand{\rhoc}{\rho_{\rm c}}
\newcommand{\hatrho}{{\hat\rho}}
\newcommand{\chim}{\chi_{\rm m}}
\newcommand{\barchim}{\bar\chi_{\rm m}}

\newcommand{\ep}{\epsilon}
\newcommand{\Rsp}{R_{\rm sp}}
\newcommand{\hsp}{h_{\rm sp}}
\newcommand{\hc}{h_{\rm c}}
\newcommand{\qc}{q_{\rm c}}
\newcommand{\Nc}{N_{\rm c}}

\makeatletter
\begin{document}

\title{A universal form of 
slow dynamics in zero-temperature random-field Ising model}

\author{Hiroki Ohta}
\email{hiroki@jiro.c.u-tokyo.ac.jp}
\author{Shin-ichi Sasa}
\email{sasa@jiro.c.u-tokyo.ac.jp}
\affiliation
%\address
{Department of Pure and Applied Sciences,
University of Tokyo, 3-8-1 Komaba Meguro-ku, Tokyo 153-8902, Japan}
\date{\today}


\begin{abstract}
The zero-temperature Glauber dynamics of the random-field Ising model 
describes various ubiquitous phenomena such as avalanches, hysteresis, 
and related critical phenomena. Here, for a model on a random graph 
with a special initial condition, we derive exactly an evolution 
equation for an order parameter.
Through a bifurcation analysis of the obtained equation,
we reveal a new class of cooperative slow dynamics
with the determination of critical exponents.
\end{abstract}

\maketitle

%% general problem / universality classes for slow dynamics

Slow dynamical behaviors caused by cooperative phenomena
are observed in various  many-body systems.
In addition to well-studied examples
such as critical slowing down \cite{H-H}, 
phase ordering kinetics \cite{Bray}, 
and slow relaxation in glassy systems \cite{Cavagna}, 
seemingly different phenomena from these examples 
have also been discovered successively. 
In order to judge  whether or not an 
observed phenomenon is qualitatively new, 
one needs to determine a universality class including 
the phenomenon. In this context, it is significant
to develop a theoretical method for classifying 
slow dynamics. 

%% bifurcation - from equilibrium to  dynamics -

Here, let us recall a standard procedure for classifying equilibrium 
critical phenomena. First,  for an order parameter $m$ of 
a mean-field model, a qualitative change in the solutions  
of  a self-consistent equation $m={\cal F}(m)$ is investigated; then, 
the differences between the results 
of the mean-field model and finite-dimensional systems 
are studied by, for example, a renormalization group 
method. On the basis of this success, an analysis 
of the dynamics of a typical mean-field model
is expected to be a first step toward determining
a universality class of slow dynamics. 

%% example of exact solvable models

As an example, 
in the fully connected Ising model with Glauber dynamics,
an evolution equation for the  magnetization,
$\partial_t m={\cal G}(m)$, can be derived exactly. The 
analysis of this equation reveals that the critical behavior 
is described by a pitchfork bifurcation in the dynamical system 
theory \cite{Gucken}. 
As another 
example, an evolution equation for a time-correlation 
function and a response function was derived exactly
for the fully connected 
spherical $p$-spin glass model \cite{CHZ, Kurchan}. The obtained 
evolution equation  represents one universality class 
related to dynamical glass transitions.

%% motivation 

The main purpose of this Letter is to present a non-trivial class
of slow dynamics by exactly deriving an evolution 
equation for an order parameter.
The model that we consider is the zero-temperature 
Glauber dynamics of a random-field Ising model,
which is a simple model for describing various ubiquitous phenomena 
such as avalanches,  hysteresis, and related critical phenomena
\cite{Inomata,Sethna,Vives,Durin,Shin1}. 
As a simple, but still non-trivial case,
we analyze the model on a random graph
\cite{fn:global}, 
which is regarded as one type of 
Bethe lattices  \cite{MP}. Thus far, 
several interesting results on the quasi-static properties of 
the model on Bethe lattices 
have been obtained 
\cite{Duxbury1,Illa,Dhar1,Colaiori1,Alava,Rosinberg0,Rosinberg1}. 
In this Letter, by performing the bifurcation analysis
of the derived equation,
we determine the critical exponents characterizing singular behaviors of 
dynamical processes.

%%%%%%%%%%%%%%%%%%%%%%%
\paragraph*{Model:}    %
%%%%%%%%%%%%%%%%%%%%%%%

% Hamiltonian and a random graph
% time-evolution rule

Let $G(c,N)$ be a regular random graph consisting of $N$ sites, 
where each site is connected to $c$ sites chosen randomly.  
For a spin variable $\sigma_i= \pm 1$ and a random field $h_i'$
on the graph $G(c,N)$, the random-field Ising model is defined by
the Hamiltonian 
\begin{equation}
H=-\frac{1}{2}\sum_{i=1}^N\sum_{j\in B_i}
\sigma_{i}\sigma_j-\sum_{i=1}^N (h+{h}_i')\sigma_i,
\label{model}
\end{equation}
where $B_i$ represents a set of sites connected to the $i$ site
and $h$ is a uniform external field. 
The random field ${h}_i'$ obeys a Gaussian distribution 
$D_R({h}_i')$ with variance $R$. 
We collectively denote $(\sigma_i)_{i=1}^N$ and $(h_i)_{i=1}^N$ by $\bv{\sigma}$ and $\bv{h}$, respectively. Let $u_i$ be the number of upward spins in $B_i$. 
Then, for a given configuration, we express the energy increment for the 
spin flip at $i$ site as $-2\sigma_i\Delta_i$, where
\begin{equation}
\Delta_i\equiv
c-2u_i-(h+{h}_i').
\label{Delta:def}
\end{equation}
The zero-temperature Glauber dynamics is defined as a stochastic
process in the limit that the temperature tends to zero for 
a standard Glauber dynamics with a finite temperature. 
Specifically, we study a case in which the initial condition is 
given by $\sigma_i=-1$ for any $i$. In this case, once $\sigma_i$
becomes positive, it never returns. Thus, 
the time evolution rule is expressed by the following simple rule: 
if $\sigma_i=-1$ and $u_i$ satisfies $\Delta_i \le 0 $, 
the spin flips at the rate of 
$1/\tau_0$; otherwise, the transition is forbidden.
Note that $\sigma_i(t)=-1$ when $\Delta_i(t) >0$, and 
$\Delta_i(t)$ is a non-increasing function of $t$
in each sample \cite{Dhar1}. In the argument below, 
a probability induced  by the probability measure 
for the stochastic time evolution for a given
realization $\bv{h}$ is denoted by  $P^{\bv{h}}$,
and the average of a quantity $X$ over $\bv{h}$ 
is represented by $\overline{X}$.

%%%%%%%%%%%%%%%%%%%%%%%%%%%%%%%%%%%%%%%%%%%%%
\paragraph*{Order parameter equation:}      %
%%%%%%%%%%%%%%%%%%%%%%%%%%%%%%%%%%%%%%%%%%%%%

% starting 

We first note that the local structure of a random graph is the same 
as a Cayley tree. In contrast to the case of Cayley trees,
a random graph is statistically homogeneous, which simplifies
the theoretical analysis. Furthermore, 
when analyzing the model on a random graph 
in the limit $N \to \infty$,  we may ignore
effects of loops. 
Even with this assumption, the theoretical analysis of dynamical 
behaviors is not straightforward, because 
$\sigma_j$ and $\sigma_k$, $j, k \in B_i$, are generally correlated. 
We overcome this difficulty by the following three-step approach.
%
% ideas
%

The first step is to consider a modified system
in which $\sigma_i=-1$ is fixed irrespective of the spin configurations.
We denote a probability in this modified system by $Q^{\bv{h}}$.
We then define $q(t)\equiv\overline{Q^{\bv{h}}(\sigma_j(t)=1)}$ for 
$j \in B_i$, where $q(t)$ is independent of $i$ and $j$ 
owing to the statistical homogeneity of the random graph.
The second step is to confirm the fact that any configurations 
with $\Delta_i(t)>0$ in the original system are realized
at time $t$ in the modified system as well, provided that 
the random field and the history of a process are identical for
the two systems. This fact leads to a non-trivial claim 
that $P^{\bv h}(\Delta_i(t) > 0)$ 
is equal to $Q^{\bv h}(\Delta_i(t) > 0)$. By utilizing 
this relation, one may express  $P^{\bv h}(\Delta_i(t) > 0)$ in terms of
$Q^{\bv{h}}(\sigma_j(t)=1)$. The average of this expression over $\bv{h}$, with the definition $\rho(t)\equiv\overline{P^{\bv h}(\Delta_i(t) > 0)}$, 
leads to 
\begin{equation}
\rho(t)
=
\sum_{u=0}^{c} 
\left(
\begin{array}{c} 
c \\ u
\end{array}
\right)
q(t)^{u}(1-q(t))^{c-u}
\int_{-\infty}^{c-2u-h} dh' D_R(h'),
\label{Gq}
\end{equation}
where we have employed the statistical independence
of $\sigma_j$ and $\sigma_k$ with $j,k \in B_i$ in the modified system.
The expression (\ref{Gq}) implies that 
$q(t)$ defined in the modified system has 
a one-to-one correspondence with the quantity $\rho(t)$
defined in the original system. 
The third step is to define 
$p(t)\equiv \overline{Q^{\bv h}(\sigma_j=-1, \Delta_j \le 0)}$
and 
$r(t)\equiv \overline{Q^{\bv h}(\Delta_j(t) >  0)}$
for $j \in B_i$.
Then, by a procedure similar to the derivation of (\ref{Gq}), 
we find that $dq(t)/dt$ is equal to $p(t)/\tau_0$.
$r(t)$ is also expressed as a function of $q(t)$ 
because $r(t)$ is equal to a probability
of $\Delta_j(t) >  0$ in a modified system 
with $\sigma_i=-1$ and $\sigma_j=-1$ fixed.
Concretely, we write $r(t)=1-F(q(t))$, where 
\begin{equation}
F(q)
=
\sum_{u=0}^{c-1} 
\left(
\begin{array}{c} 
c-1 \\ u
\end{array}
\right)
q^{u}(1-q)^{c-1-u}
\int_{c-2u-h}^{\infty} dh' D_R(h').
\label{Fq}
\end{equation}
By combining a trivial relation $p(t)+q(t)+r(t)=1$
with these results,  we obtain 
\begin{eqnarray}
\tau_0\der{q}{t}
=F(q)-q,
\label{evol:q}
\end{eqnarray}
which determines $q(t)$ with the initial condition $q(0)=0$.

% remarks

Here, for a spin configuration at time $t$
under a quenched random field $h_i'$, we define 
\begin{equation}
\hatrho(t) \equiv \frac{1}{N}\sum_{i=1}^N\theta(\Delta_i(t)),
\end{equation}
where $\theta(x)=1$ for $x > 0$ and $\theta(x)=0$ otherwise.
Due to the law of large numbers, 
$\hatrho(t)$ is equal to $\rho(t)$
in the limit $N \to \infty$. 
Since $\rho(t)$ is determined 
by (\ref{Gq}) with (\ref{evol:q}),
one may numerically compare $\rho(t)$ with $\hatrho(t)$
observed in the Monte Calro (MC) simulations. We confirmed that these
two results coincided with each other within numerical accuracy.
We also numerically checked  
the validity of (\ref{evol:q}) on the basis of
the definition of $q$. 
As a further evidence of the validity of our results, 
we remark that the stationary 
solution $q(\infty)$ satisfies $q(\infty)=F(q(\infty))$, 
which is identical to the fixed-point condition of 
a recursive equation $Z_{n+1}=F(Z_n)$ in the Cayley tree,
where $Z_n$ is a probability at generation $n$. 
(See  Ref.  \cite{Dhar1} for the precise definition of $Z_n$.)
In the argument below, we set $\tau_0=1$ without loss of generality
and we investigate the case $c=4$ as an example. 
The obtained results are essentially the same for $ c > 4$; 
however, the behaviors for $c<4$ are qualitatively 
different from those for $ c \ge 4$. 

\begin{figure}
\includegraphics[width=6.5cm,clip]{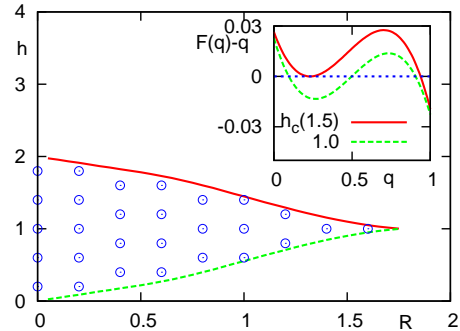}
\caption{(color online) Phase diagram. Inset: $F(q)-q$ as 
functions of $q$. $R=1.5$. $h=1.0$ and $h=h_c(R)$.}
\label{phase}
\end{figure}

%%%%%%%%%%%%%%%%%%%%%%%%%%%%%%%%%%%%%%%
\paragraph*{Bifurcation analysis:}        %
%%%%%%%%%%%%%%%%%%%%%%%%%%%%%%%%%%%%%%%

%% preliminary 

We start with the analysis of (\ref{evol:q}) for $R =1.5$ and $h =1.0$.
The qualitative behavior of $q(t)$ is understood from 
the shape of the graph $F(q)-q$ shown in the inset of Fig. \ref{phase}. 
There are three zeros, $q_1$, $q_2$, and $q_3$, 
where $0 < q_1 < q_2 <q_3 <1 $. Since $F(q) > q$ in the 
interval $[0,q_1)$, $q(t) \to q_1$ as $t \to \infty$
with the initial condition $q(0)=0$. 
%% bifurcation structure 
This geometrical structure sustains in a region of 
the parameter space $\alpha\equiv (R,h)$, as shown in 
Fig. \ref{phase}. Let $q_i(\alpha)$ be
the parameter dependence of $q_i$. Then,
the stable fixed point $q_1(\alpha)$ and the unstable 
fixed point $q_2(\alpha)$ merge at the solid line, 
and  the stable fixed point $q_3(\alpha)$ 
and the unstable fixed point $q_2(\alpha)$ merge at the dashed line,
both of which correspond to 
{\it saddle-node bifurcations} \cite{Gucken}. 
The two lines terminate at a critical point $(\Rsp, \hsp)$.
Since the trajectories with the initial condition $q(0)=0$
do not exhibit a singularity at the dashed line, 
only  the bifurcations at the solid line are 
relevant in the present problem. The solid line is 
called a spinodal line \cite{Dahmen0}.

% SN-bifurcation 

Now, we investigate the singular behaviors of slow dynamics
near the bifurcation points. We first fix the value of $R$ such 
that $0 < R \ll \Rsp$. Then, 
a saddle-node bifurcation occurs at $(R,\hc(R))$
on the solid line. Let $\qc(R)$ be the saddle-node point 
such that $q_1=q_2=\qc(R)$. We set $h=\hc(R)+\ep$ and 
$q=\qc+u$. From the graph in the inset of Fig. \ref{phase}, 
one finds that (\ref{evol:q}) becomes
\begin{equation}
\der{u}{t}=a_0\ep +a_2 u^2+O(|u|^3, |\ep u|),
\label{norm-1}
\end{equation}
when $ |u| \ll 1$ and $|\ep| \ll 1$. $a_0$ and $a_2$ are
numerical constants. 
Solutions of (\ref{norm-1}) are expressed 
as a scaling form 
\begin{equation}
u(t)=|\ep|^{1/2} \bar u_{\pm} (|\ep|^{1/2} t)
\end{equation}
when  $ |\ep| \ll 1 $, where $\bar u_{+} $ and $\bar u_{-}$
are $\ep$-independent functions for $\ep >0 $ and $\ep <0$,
respectively. The result 
implies that the characteristic time near $q=\qc$ 
diverges as 
$
\tau  \simeq |\ep|^{-1/2}.
$
Note that $q(t \to \infty)=q_3$ when $\ep >0$.
Therefore, $q(t \to \infty)$ exhibits a discontinuous change,
and the jump width is given by the distance 
between $q_1(\alpha)=q_2(\alpha)$ and $q_3(\alpha)$
at $\ep=0$.

%% CT-2 bifurcation

Next, we focus on the dynamical behaviors near 
the critical point $(\Rsp,\hsp)$. 
By substituting  $q=\qc(\Rsp)+v$ and 
$(R,h)=(\Rsp,\hsp)+(\eta,\ep)$  into (\ref{evol:q}),
we obtain 
\begin{equation}
\der{v}{t}=c_0 \ep+ c_1\eta v-c_3 v^3
+O(|v|^4, |\ep v|, |\eta v^2|),
\label{norm-2}
\end{equation}
when  $|v| \ll 1$, $|\ep| \ll 1$, and $|\eta| \ll 1$.
$c_0$, $c_1$, and $c_3$  are numerical constants.
System behaviors are classified into two types. 
First, when  $|\eta| \gg |\ep|^{2/3}$, 
solutions of (\ref{norm-2}) are expressed 
as $v=|\eta|^{1/2}\bar v_{1\pm}(|\eta| t)$
when $|\eta| \ll 1$.
This scaling form is identical to that near
a pitchfork bifurcation, which might be related 
to conjectures presented in 
Refs. \cite{Perez,Colaiori2,Dahmen1}. 
Second, when $ |\eta| \ll  |\ep|^{2/3}$, 
which includes the case in which $R=\Rsp$ is fixed, 
solutions of  (\ref{norm-2}) are expressed as 
\begin{equation}
v=|\ep|^{1/\delta}\bar v_{2 \pm}(|\ep|^{\zeta} t)
\label{scaling:1}
\end{equation}
when $ |\ep| \ll 1$, 
with $\delta=3$ and $\zeta=2/3$. 
The characteristic time diverges as 
$\tau  \simeq |\ep|^{-\zeta}$ near the critical point
$(\Rsp,\hsp)$.
In addition to the two scaling forms, one can
calculate the width of the discontinuous jump 
of $q$ along the spinodal line, which is 
proportional to $(-\eta)^{1/2}$ near the critical 
point \cite{Colaiori1,Duxbury1}.

%%%%%%%%%%%%%%%%%%%%%%%%%%%%%%%%%%%%%%%%%%%%
\paragraph*{Finite size fluctuations:}     %
%%%%%%%%%%%%%%%%%%%%%%%%%%%%%%%%%%%%%%%%%%%%

% question

In a system with large but finite  $N$, 
fluctuations of $\hat \rho$ are observed.
Their basic characterization is given by the intensity 
\begin{equation}
\chi_\rho(t)\equiv N \bra (\hatrho(t)- \bra \hatrho(t) \ket)^2 \ket.
\end{equation} where for a quantity $\hat{X}(t)$ determined by $\bv{\sigma}(t)$ and $\bv{h}$, $\bra \hat{X}(t)\ket\equiv\overline{\sum_{\bv{\sigma}} P^{\bv{h}}(\bv{\sigma}(t)=\bv{\sigma})\hat{X}(t)|_{\bv{\sigma}(t)=\bv{\sigma}}}$.
The problem here is to determine a singular behavior 
of $\chi_\rho$ under the condition that $0 < \ep  \ll 1$ and $\eta=0$. 
Let  $\rhoc$ be defined by (\ref{Gq}) with $q=\qc(\Rsp)$. 
We then assume
\begin{eqnarray}
\hatrho(t)-\rhoc=A(\ep, N)\hat{F}(t/\tau(\ep, N))
\end{eqnarray}
near $\hatrho(t) \simeq \rhoc$, where $A$ and $\tau$ are 
typical values of the amplitude and the characteristic 
time, respectively, and $\hat{F}$ is a time-dependent 
fluctuating quantity scaled with $A$ and $\tau$. 
We further conjecture  finite size scaling relations
\begin{eqnarray}
A(\ep,N) &=& N^{-1/(\nu \delta)} F_A(\ep N^{1/\nu}), \nonumber \\
\tau(\ep,N) &=& N^{\zeta/\nu}F_\tau(\ep N^{1/\nu}),
\label{tau:scale}
\end{eqnarray}
where $F_A(x) \simeq x^{1/\delta}$ for $x \gg 1$ and 
$F_\tau(x) \simeq x^{-\zeta}$ for $x \gg 1$;
$F_A(x) ={\rm const.}$ and $F_\tau(x) ={\rm const.}$ for $x \ll 1$. 
Here, the exponent $\nu$ characterizes  a cross-over size
$\Nc$ between the two regimes as a power-law form 
$\Nc \simeq \ep^{-\nu}$. We thus obtain 
\begin{equation}
\chi_\rho(\tau(\ep,N))=N^{\gamma/\nu}F_\chi(\ep N^{1/\nu}),
\label{fluc:scale}
\end{equation}
where $\gamma=\nu-2/\delta$. The values of $\zeta$
and $\delta$ have already been determined. 
We derive the value of $\nu$ in the following
paragraph.

% determination of \nu

It is reasonable to assume that the qualitative behavior 
of $\hat \rho$ near the critical point $(\Rsp,\hsp)$ is 
described by (\ref{norm-2}) with small fluctuation effects
due to the finite size nature.
There are two types of fluctuation effects: one 
from the stochastic time evolution and the other 
from the randomness of $h_i'$. The former
type is expressed by the addition of a weak Gaussian-white 
noise with a noise intensity proportional to $1/N$,
whereas the latter yields  a weak quenched disorder of 
the coefficients of  (\ref{norm-2}). In particular,  
$\ep$ is replaced with $\ep+ \hat g/\sqrt{N}$, where 
$\hat g$ is a time-independent quantity that obeys a 
Gaussian distribution. Then, two characteristic sizes 
are defined  by a balance between the fluctuation effects
and the deterministic driving force. As the influence of
the stochastic time evolution rule, the size $N_{\rm s}$
is estimated from a dynamical action of the path-integral 
expression 
$\int dt[ N c_4 (\partial_t v-c_{0} \ep +c_3 v^3)^2+c_5 v^2]$,
where the last term is a so-called Jacobian term. 
$c_4$ and $c_5$ are constants. In fact, the balance among the  
terms $\ep^2 \simeq v^6 \simeq v^2/N_{\rm s}$ leads to 
$N_{\rm s} \simeq \ep^{-4/3}$. (See Refs. \cite{Iwata,Ohta2} for
a similar argument.) On the other hand, the size $N_{\rm q}$
associated with the quenched disorder is determined by the 
balance $\ep \simeq 1/\sqrt{N_{\rm q}}$,
which leads to $N_{\rm q} \simeq \ep^{-2}$. 
Since $N_{\rm s} \ll N_{\rm q}$ for $\ep \ll 1$,
the system  is dominated by fluctuations 
when $N \le N_{\rm q}$.
With this consideration, we conjecture that 
$\Nc=N_{\rm q} \simeq \ep^{-2}$. That is, $\nu=2$.

% numerical confirmation 

\begin{center}
\begin{figure}
\includegraphics[width=6.5cm]{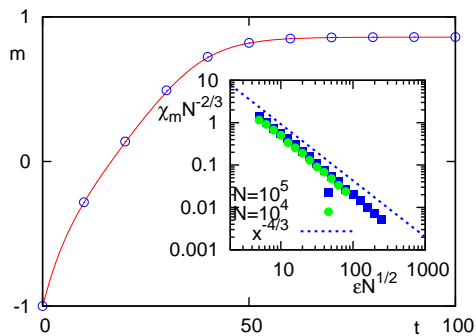}
\caption{(color online) 
Time evolution of the magnetization for $R=1.781258(\simeq \Rsp)$ and $h=1.1$. 
Note that $\hsp=1$. The circles are the result obtained by the MC simulations with $N=10^5$,
and the solid line corresponds to the solution of 
$\partial_t m=-(m-1+2\rho)$ 
with (\ref{Gq}) and (\ref{evol:q}). One may derive this equation 
for $m$ in a straightforward manner. 
Inset: $\chim N^{-2/3}$ as functions of $\ep N^{1/2}$. 
The guideline represents $\chim \simeq \ep^{-4/3}$.
}
\label{scale}
\end{figure}
\end{center}

In laboratory and numerical experiments, 
statistical quantities related to the magnetization 
$\hat m= \sum_{i=1}^N \sigma_i/N$ may be measured
more easily than $\chi_\rho$.
Since $\hat m$ is not independent of $\hatrho$, 
$\hat m$ also exhibits singular behaviors 
near the critical point.  
Concretely, the fluctuation intensity
$\chi(t)\equiv N \left(\bra \hat{m}^2(t) \ket
-\bra \hat{m}(t)\ket^2\right)$ is characterized
by the above exponents. In order to confirm
this claim, we measured $\chi(t)$ by MC simulations.
Then, the characteristic time is defined as a time $t_*$
when $\chi(t)$ takes a maximum value $\chim$. 
Our theory predicts $t_* \simeq \ep^{-2/3}$ and 
$\chim=N^{2/3} \barchim(\ep N^{1/2})$ with 
$\barchim(x) \simeq x^{-4/3}$ for $x \gg 1$.
The numerical results shown in Fig. \ref{scale}
are consistent with the theoretical predictions.

%%%%%%%%%%%%%%%%%%%%%%%%%%%%%%%%%%%%%%%%%%%%%%%%%%%%%%%%%%
\paragraph*{Concluding remarks:}\label{Remark}            %
%%%%%%%%%%%%%%%%%%%%%%%%%%%%%%%%%%%%%%%%%%%%%%%%%%%%%%%%%%

We have derived the exact evolution equation for the order
parameter $\rho$ describing the dynamical behaviors of a
random field Ising model on a random graph. From this 
dynamical system, we 
have determined the values of the critical exponents:
$\zeta=2/3$, $\delta=3$, $\nu=2$, and  $\gamma=4/3$. 

% upper-critical dimension

Before ending this Letter, we discuss the critical 
phenomena in $d$-dimensional random-field Ising models
on the basis of our theoretical results.
 Let $\nu_{\rm m}$ be the critical exponent 
characterizing the divergence of a correlation length above an 
upper-critical dimension $d_{\rm u}$. By assuming a standard 
diffusion coupling for (\ref{norm-2}) as 
an effective description of finite-dimensional 
systems, we expect $z\equiv\zeta/\nu_{\rm m}=2$. This leads 
to $\nu_{\rm m}=1/3$. From an application of a hyper-scaling
relation to the upper-critical dimension, one also expects 
the relation $\nu=d_{\rm u}\nu_{\rm m}$ \cite{Pfeuty}.
This leads to $d_{\rm u}=\nu/\nu_{\rm m}=6$, which is consistent
with the previous result \cite{Dahmen0}.  We then denote the 
exponents characterizing the divergences of time and length 
scales by $\nu_{3}$ and $\zeta_{3}$, respectively, for 
three-dimensional systems. These values were estimated
as $\zeta_3\simeq 1.3$ and $\nu_3\simeq 0.7$ in numerical
experiments \cite{Carpenter,Perez,Ohta1}. 
The theoretical analysis of finite dimensional systems will be attempted as 
an extension of the present work; this might be complementary to previous studies \cite{Dahmen0,Muller}. Finally, we wish to mention that
one of the most stimulating studies is to 
discover such a universality class in laboratory experiments
by means of experimental techniques that capture spatial 
structures directly \cite{Shin1,Shin2,Durin}. 

% acknowledgment

We acknowledge M. L. Rosinberg, G. Semerjian, 
S. N. Majumdar, H. Tasaki, and G. Tarjus 
for the related discussions. 
This work was supported by a grant from 
the Ministry of Education, Science, Sports and Culture of Japan, 
Nos. 19540394 and 21015005. 
One of the authors (H.O.) is supported by a Grant-in-Aid for 
JSPS Fellows.

\end{document}